\documentclass{aa}
\usepackage{setspace}
\usepackage{graphicx}
\usepackage{color}
\usepackage{float}
\usepackage{amsmath}
\usepackage{threeparttable}
\usepackage{txfonts}
\usepackage{natbib}
\usepackage{subfigure}
\usepackage{gensymb}
\usepackage{bm}
\usepackage{caption}
\usepackage{overpic}
\begin{document} 
\title{Investigation of lunar ejecta dynamics: Particles reaching the near-Earth space and their effect on Earth-based observation}
\author
{Kun Yang\inst{1} 
\and
 Yu Jiang\inst{2}
\and
 Youpeng Liang\inst{1}
\and
 Xiaodong Liu\inst{1,3}}
\institute
{ School of Aeronautics and Astronautics, Shenzhen Campus of Sun Yat-sen University, 518107 Shenzhen, China 
\and
 State Key Laboratory of Astronautic Dynamics, Xi'an Satellite Control Centre, 710043 Xi'an, China
\and
 Shenzhen Key Laboratory of Intelligent Microsatellite Constellation, Shenzhen Campus of Sun Yat-sen University, 518107 Shenzhen, China\\
\email{liuxd36@mail.sysu.edu.cn}}

\abstract
{}
{Particles ejected from the lunar surface via hypervelocity impacts form a torus between the Earth and the Moon. According to our previous study (Yang et al., A\&A, 659, A120), among them about $2.3\times10^{-4}\,\mathrm{kg/s}$ particles impact the Earth after long-term orbital evolution.
We mainly focus on these Earth impactors, analyze their orbital element distribution, and estimate their influence on Earth-based observations.} 
{In previous work we simulated the long-term orbital evolution of particles ejected from the lunar surface, and obtained their steady-state spatial distribution in the Earth-Moon system. In this work, we analyze the simulation results about the Earth impactors, including the fraction of impactors with different initial parameters among all impactors, the orbital element distribution, and the projection of particles onto several Earth-based observatories.}
{Particles ejected from the lunar surface are more likely to impact the Earth within a certain range of initial parameters. Most of these lunar-ejected impactors ($\sim70\%$) reach the Earth within one year, while most of the small ones ($87.2\%$ of $0.2\,\mathrm{\mu m}$ particles and $64.6\%$ of $0.5\,\mathrm{\mu m}$ particles) reach the Earth within one week. A large proportion of lunar-ejected Earth impactors can be distinguished from interplanetary dust particles according to the differences in their orbital distributions. Besides, lunar-ejected particles may exhibit distinct configurations and orientations from the perspectives of different Earth-based observatories.} 
{}
\titlerunning{Investigation of lunar ejecta dynamics}
\authorrunning{Yang \& Liu.} 
\maketitle 
\section{Introduction}

A "horizon glow" engulfing the Moon was first photographed by Surveyor 5, which was attributed to the scattering of sunlight by charged lunar dust \citep{rennilson1974surveyor}. Since then, several probes (e.g., Aollo 15 and 17) indirectly detected the possible dust in the vicinity above the lunar surface \citep{o2011review}. In 1990, the Munich Dust Counter (MDC) onboard HITEN spacecraft recorded approximately 150 dust impacts on the lunar orbit with perilune between $100$-$8000\,\mathrm{km}$ and apolune of about $50,000\,\mathrm{km}$, but failed to detect particles near the lunar surface due to its high orbit \citep{iglseder1996cosmic}. Later, the Lyman-Alpha Mapping Project (LAMP) far-ultraviolet spectrograph carried by the Lunar Reconnaissance Orbiter (LRO) aimed to search for the reported "horizon glow" but unfortunately did not observe any dust scattering phenomenon \citep{feldman2014upper}. Until 2013, the Lunar Dust EXperiment (LDEX) onboard the Lunar Atmosphere and Dust Environment Explorer (LADEE) detected an asymmetric and permanent dust cloud around the Moon \citep{horanyi2015permanent}. 
   
Based on the LDEX measurements, many efforts have been devoted to studying the nature of the lunar dust cloud. \cite{horanyi2015permanent} obtained the size distribution of dust particles, and found the variation of dust density with their altitude and also the local time. \cite{szalay2019impact} focused on the impact-ejecta model of particles in the lunar polar regions and derived the dust production rate considering six previously identified meteoroid populations. Later the model was improved by \cite{pokorny2019meteoroids} via introducing the nightside/dayside imbalance of lunar dust production. Previous studies on the lunar dust dynamics mainly focused on the source of impactors, the production model and the electrostatic migration of bound particles \citep[e.g.,][]{jones1993sporadic, pokorny2019meteoroids, hartzell2013dynamics}. However, fewer attentions have been paid to particles that can escape the Moon (hereafter referred as "escapers"). Although most of dust particles ejected from the lunar surface via hypervelocity impacts eventually fall back \citep{Szalay2016Lunar}, a small proportion ($\sim 2\%$) of them can reach the farther Earth-Moon space, and form a steady-state dust torus with a peak number density on the order of $10^{-9}\,\mathrm{m^{-3}}$ \citep{yang2022}, which is comparable to that of interplanetary dust at 1~AU ($\sim10^{-10}\,\mathrm{m^{-3}}$ \citep{grun1985collisional,dikarev2005new}). Particles may cause the negative effect on the astronomical observation for Earth-based observatories \citep{fladeland2022investigations}. 

Several studies have been devoted to analyzing the dynamical transfer of lunar ejecta. \cite{gault1983terrestrial} found that $0.5\%$ of lunar-ejected particles ultimately impacted the Earth, and the accretion rate of lunar ejecta onto the Earth is about $10^7$-$10^8\,\mathrm{g/yr}$, corresponding to $10^{-4}$-$10^{-3}\,\mathrm{kg/s}$. \cite{alexander1984mass} analyzed the transport of lunar ejecta in the Earth-Moon system, and inferred that lunar ejecta flux at the Earth's magnetopause depended on the lunar phase angle. \cite{gladman1995dynamical} found that $23\%$-$50\%$ lunar-ejected particles were re-accreted by the Earth within 10 Myrs. \cite{yamamoto1996orbital} estimated the average speed of particles ($\sim10\,\mathrm{km/s}$) when they reached the near-Earth region, and concluded that the impact angles of particles were nearly isotropic. \cite{fritz2012impact} calculated the total mass of lunar ejecta transferred to the Earth in $10^6$ years, which was about $2.0\times10^{-4}\,\mathrm{kg/s}$. \cite{fladeland2022investigations} projected the lunar-ejected particles to the sky of an Earth-based observatory located in the northern hemisphere, and suggested that particles were visible almost every night, with distinct orientations and configurations.

This work aims to investigate the particles ejected from the lunar surface and finally impacting the Earth (hereafter referred as "Earth impactors"), analyze their orbital characteristics, and estimate their effects on the Earth-based observation. The paper is organized as follows. In Sect.2 the dust dynamical model and the simulation scheme shown in our previous paper are presented briefly. Besides, several coordinate systems used in our simulation are introduced. In Sect.3 simulation results about particles impacting the Earth are presented, including the fraction of impactors with different initial parameters among all impactors, the distribution of orbital elements at the reaching moment, and the orbital evolution for an Earth impactor. In Sect.4 lunar-ejected particles are projected to several noted observatories located at different regions of the Earth, to analyze their effect on the Earth-based astronomical observation.
\label{sect1}

\section{Method}
\subsection{Dust dynamical model and simulation scheme}
In our previous work, we have established the dynamical model and performed the long-term orbital simulations for particles ejected from the lunar surface. Here, we briefly introduce the dynamical model and the simulation scheme.

The initial parameters of particles ejected from the lunar surface for the simulations are presented in Table \ref{tab:1}. Both the size and the initial velocity of particles follow the power-law distribution, with exponents of $3.7$ and $2.2$, respectively. To eliminate the seasonal effect, twelve starting moments for particles were chosen, covering one period of the lunar ascending node precession ($18.6$ Earth years). Besides, particles were assumed to be ejected vertically at $200$ starting regions of the lunar surface, corresponding to ten ejection latitudes and twenty ejection longitudes. The gravity of Earth, solar radiation pressure, Poynting-Robertson drag, and gravity perturbations of the Sun and the Moon were considered in the dynamical model. In total, the motions of $216,000$ particles ejected from the lunar surface were simulated until they hit the Moon, hit the Earth or escape from the Earth-Moon system, with the maximum motion time ($100$ Earth years). For more details about the dynamical modeling and the simulation scheme, readers are referred to \cite{yang2022}.

\begin{table}[htbp]
    \centering
    \caption{Assumed initial parameters of lunar-ejected particles for simulations. The values of the bulk density and the mass production rate for lunar-ejected particles are taken from \cite{solomon1974density} and \cite{pokorny2019meteoroids}, respectively. The variables $r_\mathrm{min}$ and $r_\mathrm{max}$ are the minimum and maximum radii of lunar-ejected particles, respectively. The variables $v_\mathrm{min}$ and $v_\mathrm{max}$ are the minimum and maximum ejection velocities, respectively, the values of which are taken from the measurement of the LDEX \citep{horanyi2015permanent}, and $v_{\mathrm{esc}}$ is the escape velocity from the lunar surface.}
    \begin{tabular}{cc}
        \hline\hline
         Parameter& Value \\ 
         \hline
         bulk density& $3500\,\mathrm{kg/m^3}$\\
         mass production rate& $0.2\,\mathrm{kg/s}$ \\
         $r_\mathrm{min}$ & $0.2\,\mathrm{\mu m}$\\
         $r_\mathrm{max}$ & $100\,\mathrm{\mu m}$\\
         $v_\mathrm{min}$&$0.95\,v_\mathrm{esc}$\\
         $v_\mathrm{max}$&$2.00\,v_\mathrm{esc}$\\
         \hline
    \end{tabular}  
    \label{tab:1}
\end{table}

\subsection{Coordinate systems used in our work}

The integration of particles' motion is performed in the J2000 Earth-centered inertial (ECI) frame. Unless otherwise specified, the view angles of particles are provided in the "east, north, up" (ENU) local tangent plane frame. The coordinates of particles are firstly transformed from the ECI frame to the Earth-centered, Earth-fixed (ECEF) frame using the "sxform\_c" function from the NAIF SPICE toolkit \footnote{https://naif.jpl.nasa.gov/naif/credit.html}, and then the ECEF coordinates are converted to the ENU coordinates \citep[e.g.][]{meeus1991astronomical}.

\section{Particles reaching the Earth}
\subsection{Effect of initial parameters}
Via averaging the initial parameters (including the ejection longitude, the latitude, the velocity, and the particle size), the mass rate of Earth impactors has been estimated to be $2.3\times10^{-4}\,\mathrm{kg/s}$ \citep{yang2022}, which is comparable to the values reported by previous studies, i.e.~$10^{-4}$-$10^{-3}\,\mathrm{kg/s}$ by \cite{gault1983terrestrial} and $2.0\times10^{-4}\,\mathrm{kg/s}$ by \cite{fritz2012impact}. In this work we analyze the simulation results about these lunar-ejected Earth impactors ($8,077$ particles in total).

\begin{figure}[htbp]
    \centering    \includegraphics[width=0.45\textwidth]{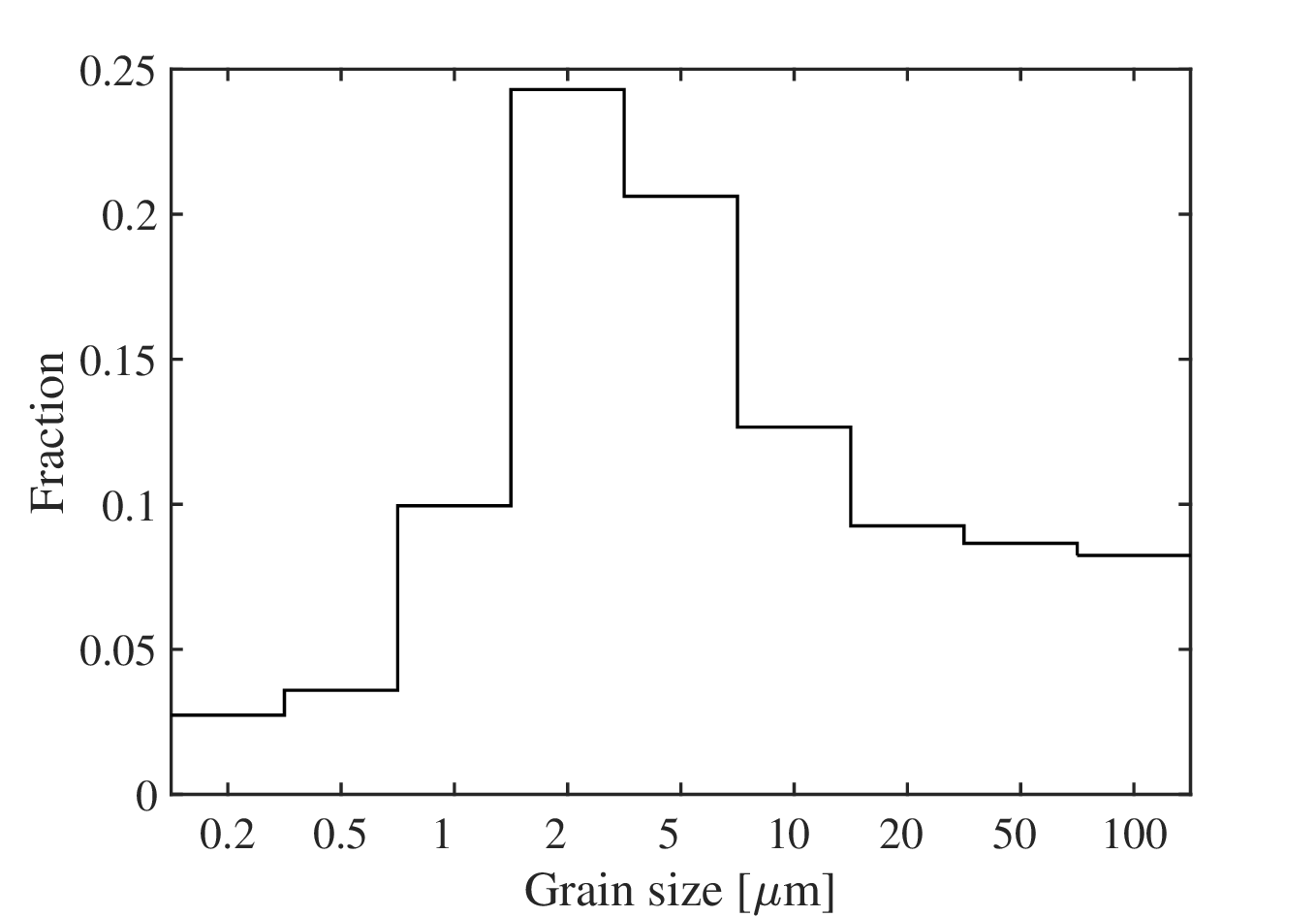} 
    \caption{Fraction of impactors vs.~grain size.} 
    \label{fig:fraction_size}
\end{figure}
% 删掉(a).?
% 已删
Fig.~\ref{fig:fraction_size} depicts the fraction of Earth impactors with different sizes. The fraction shows a rapid increase in the size range of $[0.2\,\mathrm{\mu m},2\,\mathrm{\mu m}]$, then drops to approximately $3.6\%$ for particles with the size of $100\,\mathrm{\mu m}$ (Fig.~\ref{fig:fraction_size}). This can be attributed to the size-dependent solar radiation pressure, which can blow out most of the small particles ($\leq 0.5\,\mathrm{\mu m}$) and stay most of the large particles ($\geq10\,\mathrm{\mu m}$) in the dust torus between the Earth and the Moon. 

\begin{figure}[H]
    \centering
    \includegraphics[width=0.45\textwidth]{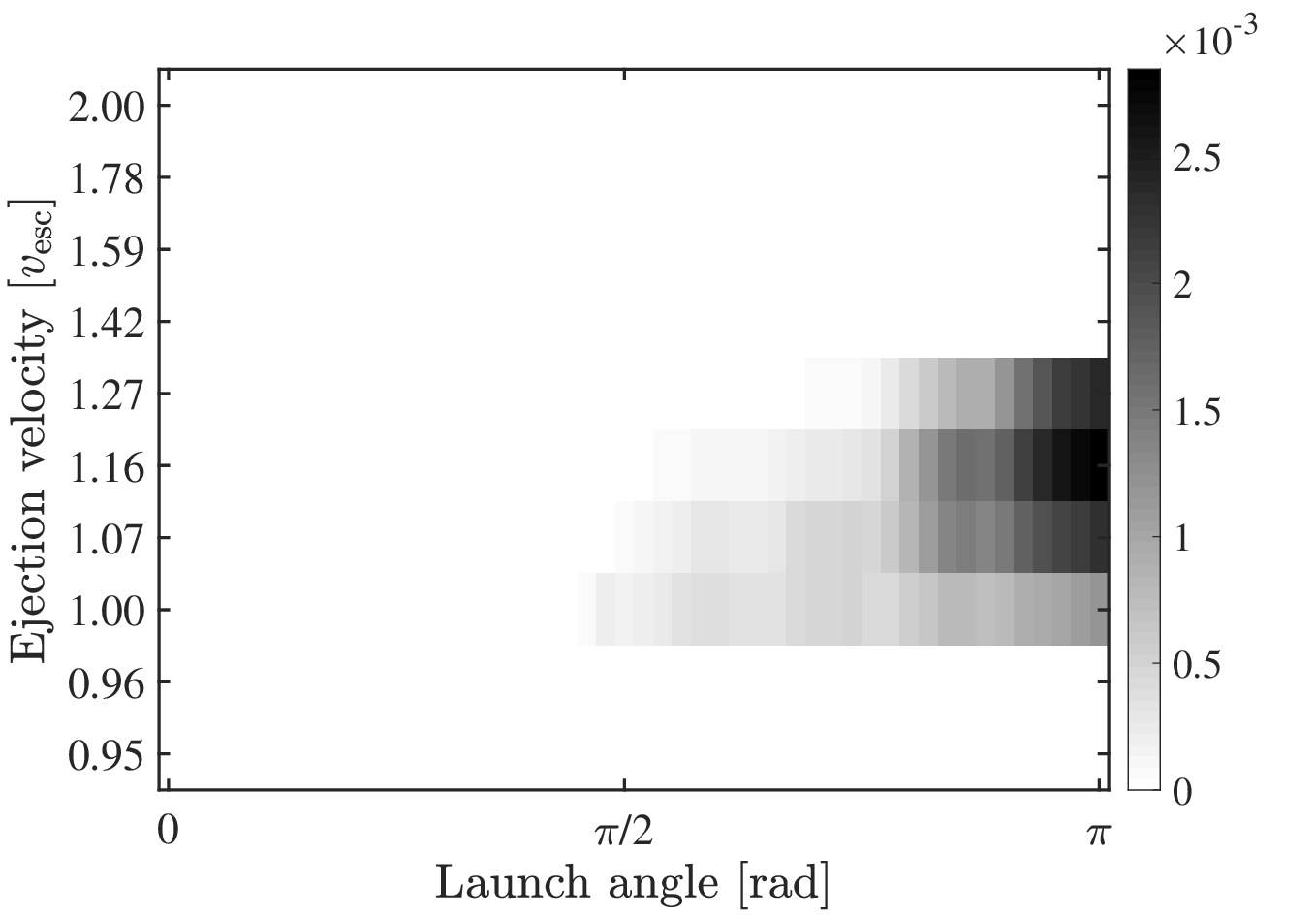}
    \caption{Fraction of impactors with different launch angles and ejection velocities.}
    \label{fig:lav}
\end{figure}

In addition to the grain size, the launch angle (the angle between the ejection velocity of the particle and the motion direction of the Moon \citep{gladman1995dynamical}) and ejection velocity also affect the orbital evolution and the transfer probability of lunar-ejected particles. From Fig.~\ref{fig:lav}, only particles with ejection velocities in the range of $\left[v_\mathrm{esc},1.27\,v_\mathrm{esc}\right]$ have the chance to impact the Earth, which is reasonable since the low-velocity ($<v_\mathrm{esc}$) particles are unable to escape the lunar gravity, and particles with ejection velocities $\geq 1.42\,v_\mathrm{esc}$ tend to evolve into hyperbolic orbits and leave the Earth-Moon system rapidly. As for the launch angle, particles with launch angles $>\pi/2$ are likely to transfer from the Moon to the Earth, with the highest transfer probability occurring at the launch angle of $\pi$, i.e.~when the particles are ejected in the opposite direction of the Moon's motion around the Earth. Particles with launch angles $<\pi/2$ can hardly reach the Earth, which is consistent with \cite{gladman1995dynamical}.

\subsection{Orbital distribution and evolution}

\begin{figure}[htbp]
    \centering
    \subfigure[]
    {    \includegraphics[width=0.45\textwidth]{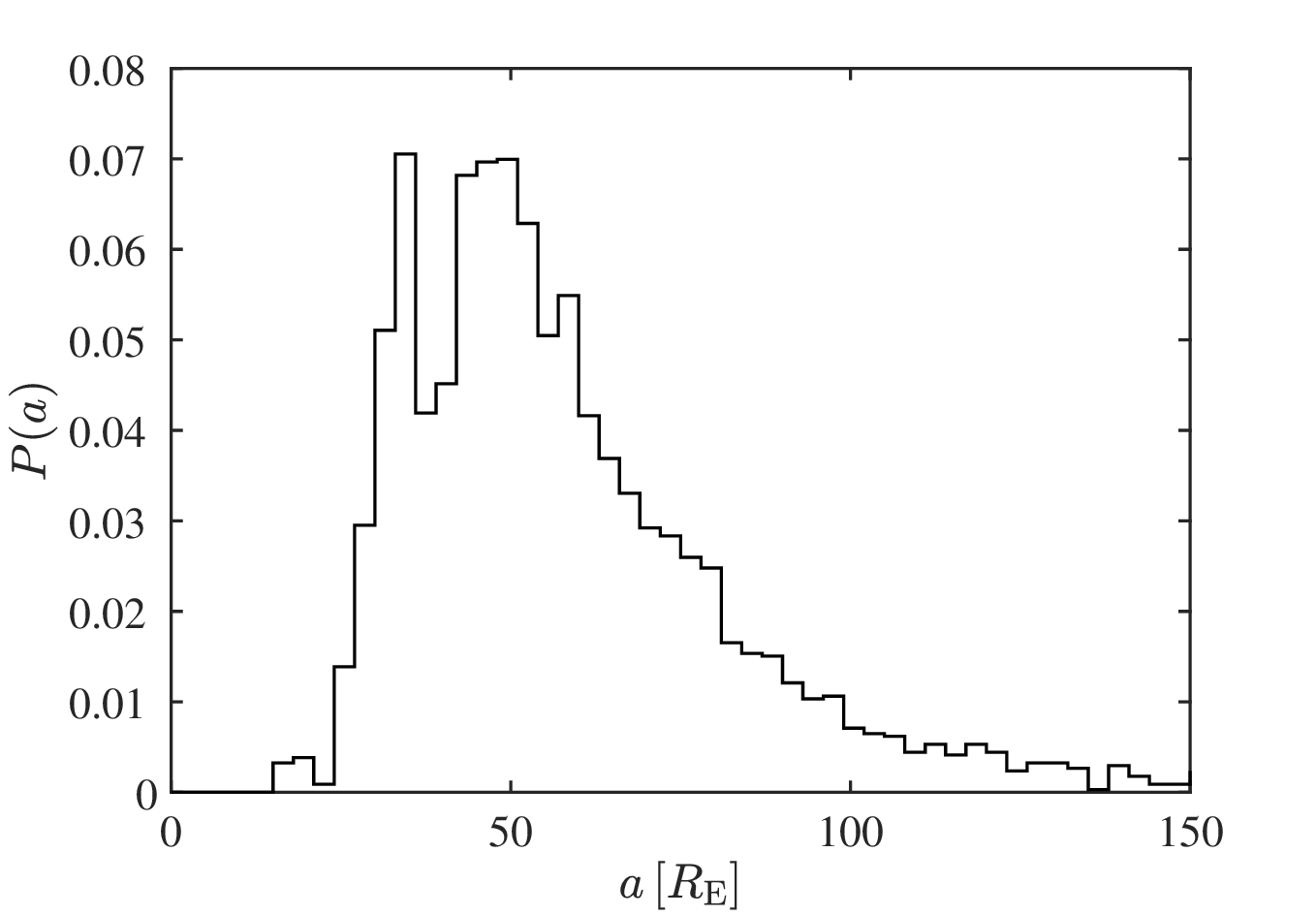}
    \label{fig:reaching_a}
    }
    \subfigure[]
    {    \includegraphics[width=0.45\textwidth]{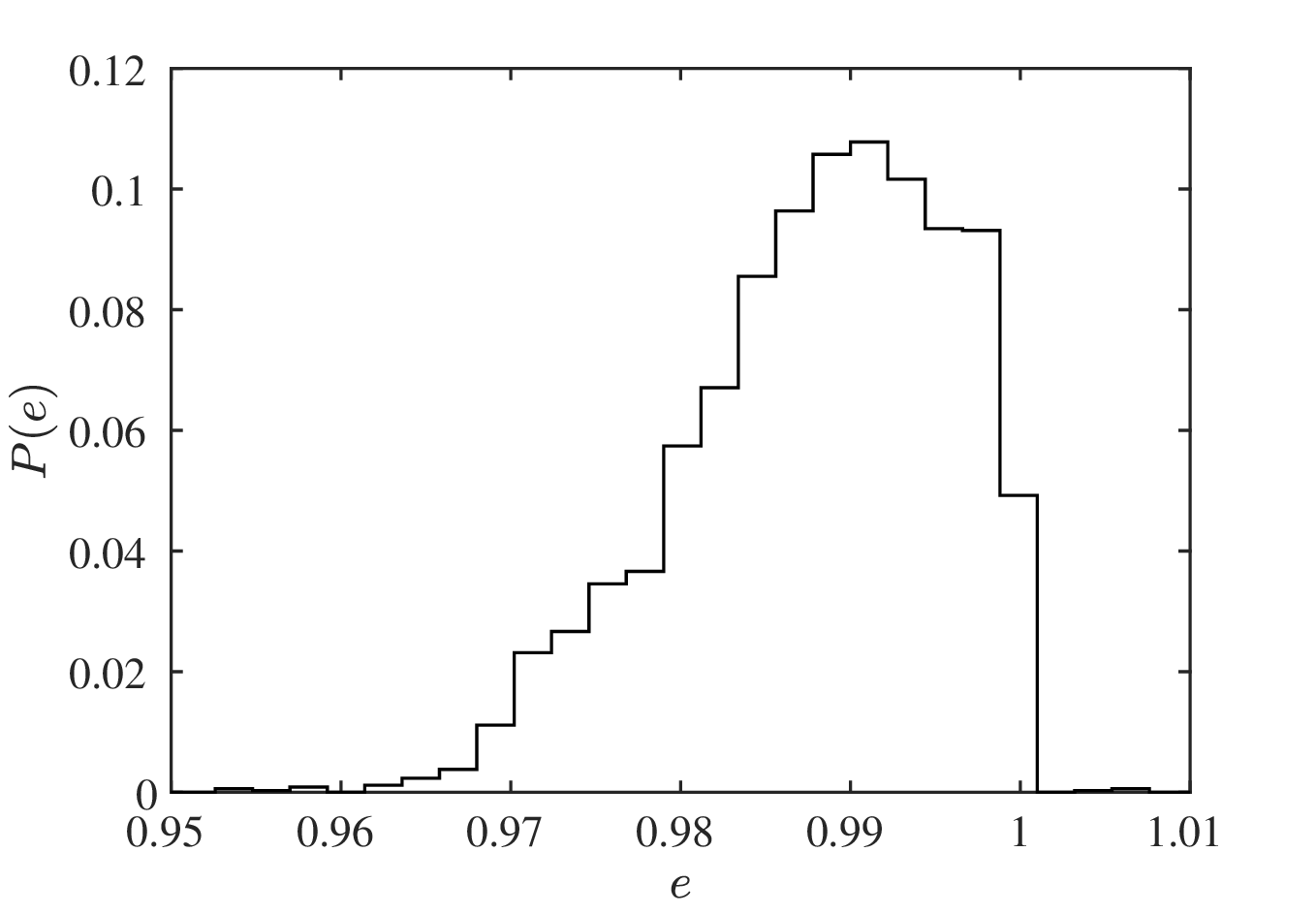}
    \label{fig:reaching_e}
    }
    \subfigure[]
    {    \includegraphics[width=0.45\textwidth]{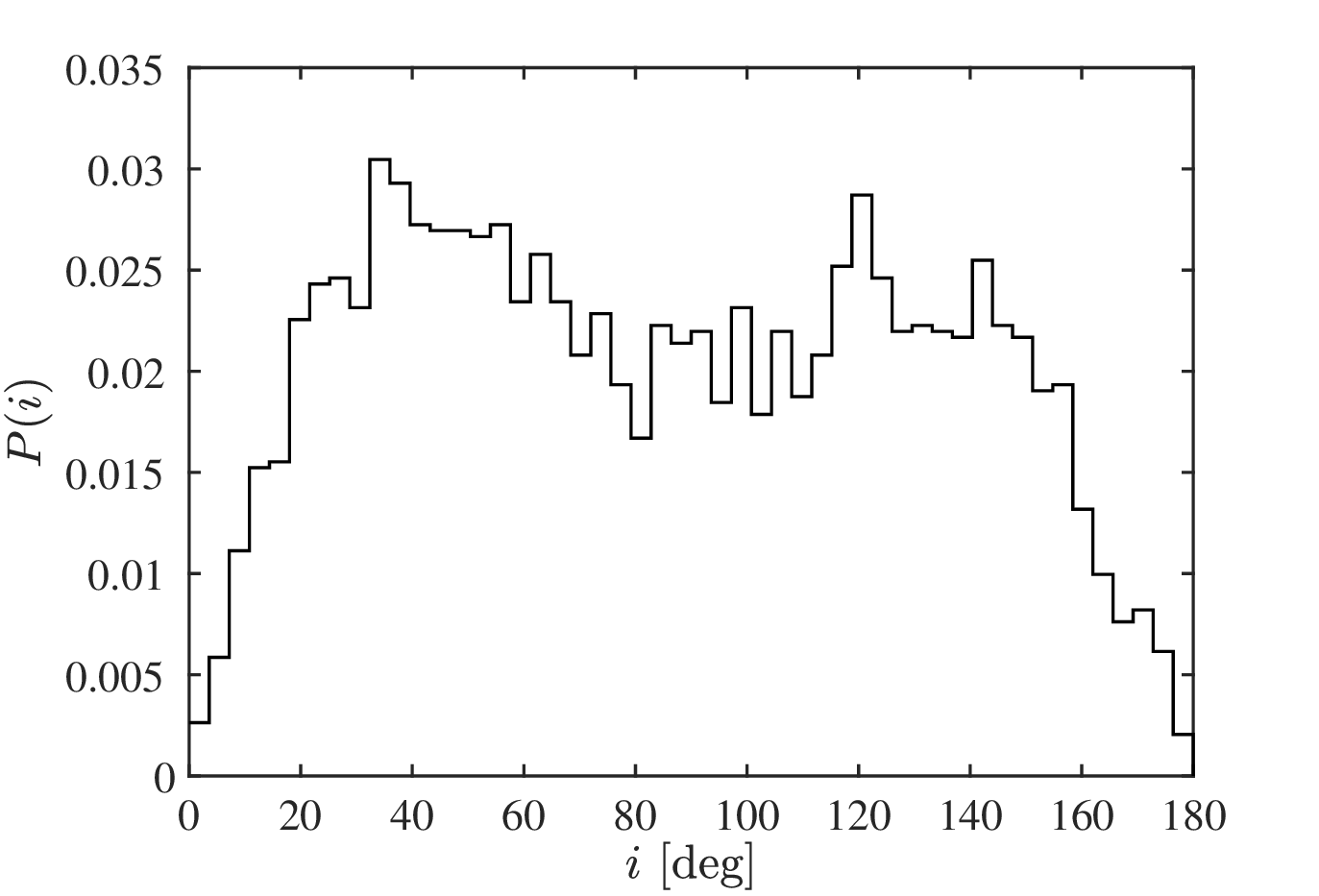}
    \label{fig:reaching_i}
    }
    \caption{Distribution of $a$, $e$ and $i$ for lunar-ejected particles at the moment of reaching the Earth. The symbols of $a$, $e$ and $i$ denote the semi-major axis, the eccentricity and the inclination of particles, which are measured in the ECI frame. \emph{Panel a}: Distribution of $a$. \emph{Panel b}: Distribution of $e$. \emph{Panel c}: Distribution of $i$.} 
    \label{fig:reaching_element}
\end{figure}

The orbital elements of lunar-ejected particles at the moment of reaching the Earth are shown in Fig.~\ref{fig:reaching_element}. Note that Fig.~\ref{fig:reaching_element} only includes particles that never leave the Earth’s Hill sphere. The semi-major axes of particles are mainly distributed in the range of $[20, \, 150]\,R_\mathrm{E}$, with a peak at about $50\,R_\mathrm{E}$. The eccentricities of particles are in the range of [0.96, 1], peaking at about 0.99. The inclinations of lunar-ejected particles are dispersed in the range of $\left[0^\circ,\,180^\circ\right]$. Among them, about $54.0\%$ particles are on prograde orbits, peaking near $40^\circ$, while a large number ($\sim46.0\%$) of particles are on retrograde orbits, peaking near $120^\circ$.
% The instantaneous positions and velocities of dust particles near the Earth can be recorded by in-situ dust detector on board spacecraft, and then transformed into orbital elements. Therefore, a large proportion of lunar-ejected particles can be distinguished from interplanetary dust particles near the Earth, especially according to the differences in their inclination distributions.

\begin{figure}[htbp]
    \centering
    \includegraphics[width=0.45\textwidth]{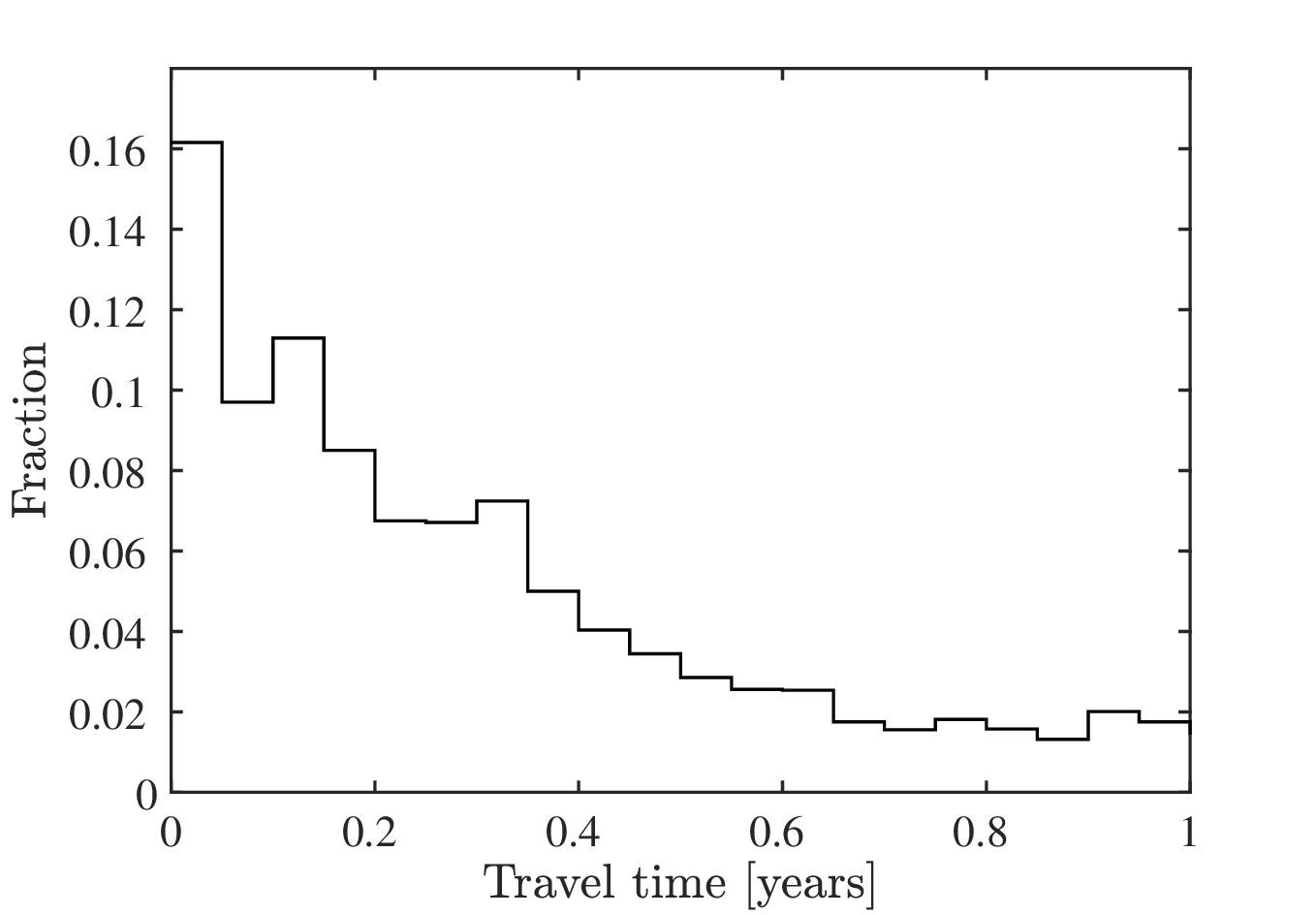}
    \caption{Distribution of travel time of Earth impactors.}
    \label{fig:travel_time}
\end{figure}

The distribution of travel time for lunar-ejected Earth impactors is shown in Fig.~\ref{fig:travel_time}. Only particles with travel times less than one year are shown in this figure, which account for $\sim70.0\%$ of all particles in our simulation. Notably, the travel time in our simulations is longer than those reported by previous studies \citep[e.g.,][]{alexander1984mass,yamamoto1996orbital}. It may 
attribute to different grain sizes selected for simulations between previous studies and this paper ($\leq 0.6\,\mathrm{\mathrm{\mu m}}$ in \citealt{alexander1984mass}; $<1\,\mathrm{\mu m}$ in \citealt{yamamoto1996orbital}; $0.2$-$100\,\mathrm{\mu m}$ in this paper).
In our simulations, we note that most of the small impactors ($87.2\%$ of $0.2\,\mathrm{\mu m}$ particles and $64.6\%$ of $0.5\,\mathrm{\mu m}$ particles) transfer from the Moon to the Earth within one week, which is consistent with previous studies.

\begin{figure}[htbp]
    \centering
    \subfigure[]
    {    \includegraphics[width=0.45\textwidth]{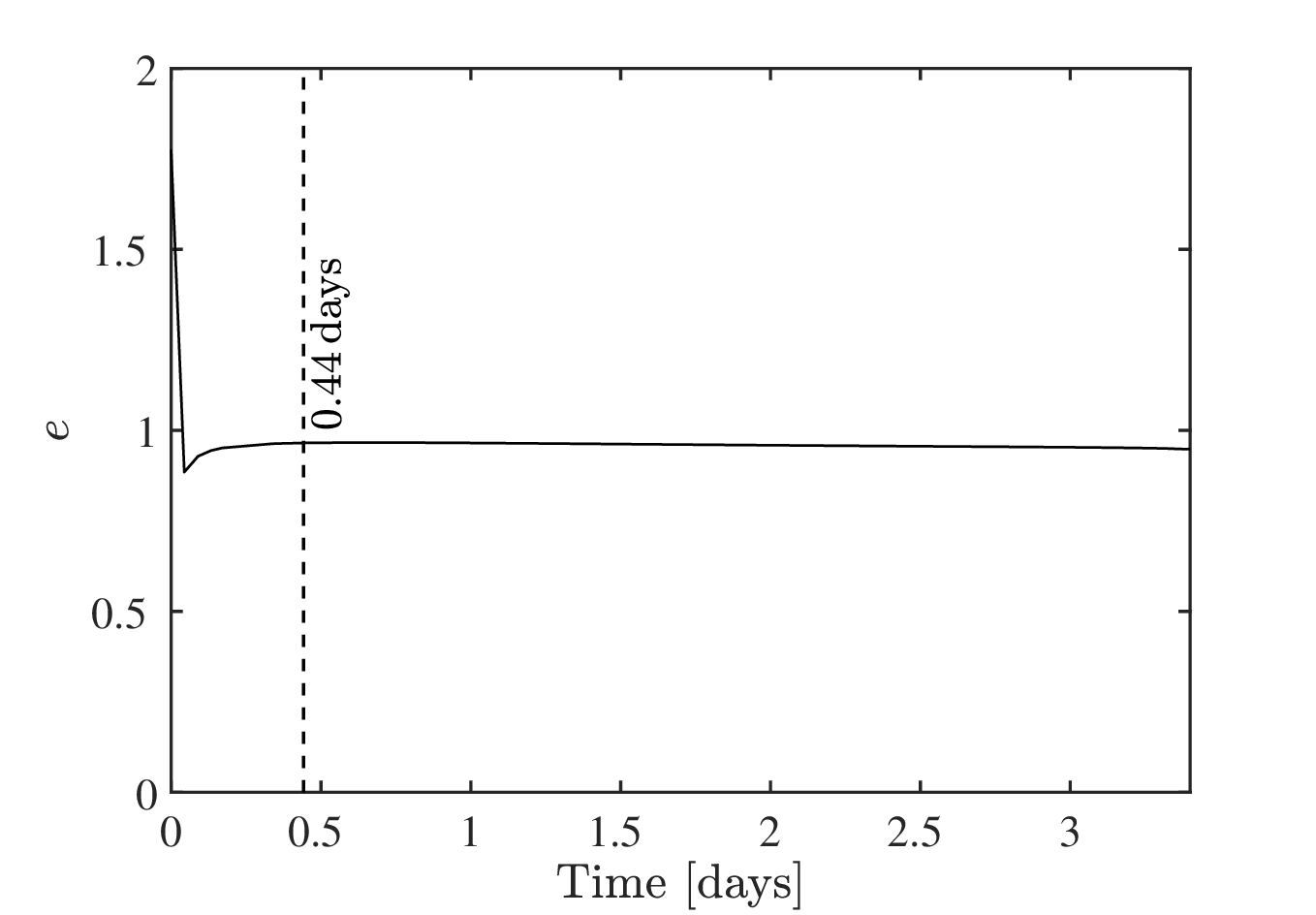}
    \label{fig:evo_e}
    }
    \subfigure[]
    {    
    \includegraphics[width=0.45\textwidth]{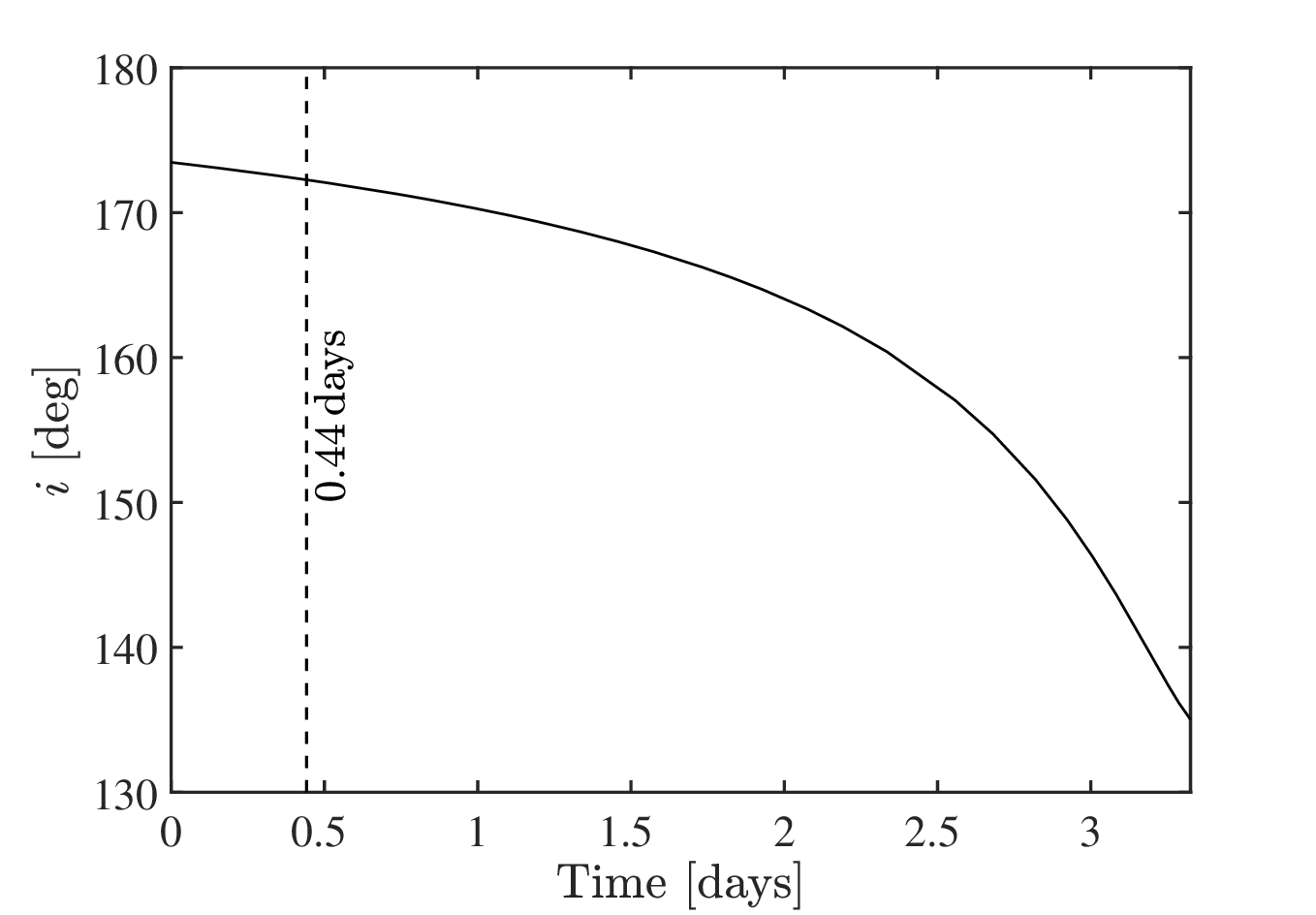}
    \label{fig:evo_i}
    }
    \subfigure[]
    {    \includegraphics[width=0.45\textwidth]{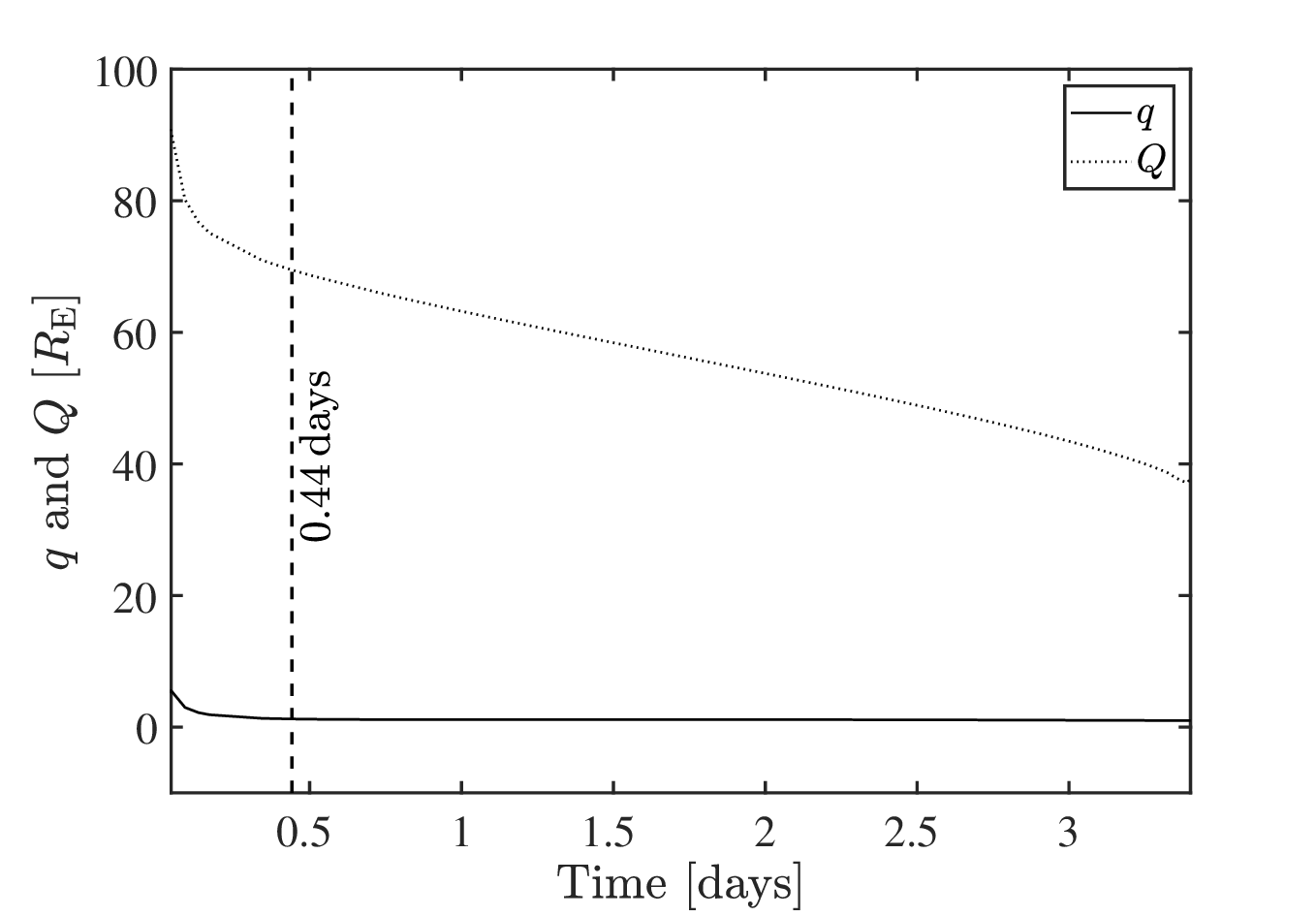}
    \label{fig:evo_qQ}
    }
    \caption{Evolution of $e$, $i$, $q$ and $Q$ for a $0.2\,\mathrm{\mu m}$ impactor from initial retrograde orbit, where $e$, $i$, $q$ and $Q$ denote the eccentricity, the inclination, the perigee and the apogee of the particle, which are measured in the geocentric lunar orbital frame. The dashed lines denote the time at which the particle leaves the Moon's Hill sphere. \emph{Panel a}: Evolution of $e$. \emph{Panel b}: Evolution of $i$. \emph{Panel c}: Evolution of $q$ and $Q$. }
    \label{fig:evo_element}
\end{figure}

Fig.~\ref{fig:evo_element} presents the evolution of orbital elements for a $0.2\,\mathrm{\mu m}$ particle from an initial retrograde orbit. The particle is ejected from the lunar surface with a speed of 2.71 km/s, and then decelerates as it climbs out of the Moon's gravitational well. As it exits the Moon's Hill sphere, its geocentric eccentricity is 0.97 (see Fig.~\ref{fig:evo_e}). Then, the particle moves on highly elliptical orbits until it impacts the Earth after about 3.4 days. From Fig.~\ref{fig:evo_i}, this particle always moves on retrograde orbits during the whole transfer process, and its inclination decreases continuously from the initial $173.5^\circ$ to $134.2^\circ$. The perigee of this particle decreases continuously, while the apogee remains nearly constant around one Earth radius after a rapid descent at the beginning (Fig.~\ref{fig:evo_qQ}).

\section{Projection onto Earth-based observatory}

In order to estimate the effect of lunar-ejected particles on the Earth-based observation, the projection of particles onto several observatories can be obtained by the transformation from the ECI frame to the ENU topocentric frame. Five observatories are selected in this work, the locations of which are listed in Table~\ref{tab:2}. Lunar-ejected particles are visible (above the local horizon) for all selected observatories, forming dust belts with different orientations and configurations.

\begin{table}[htbp]
    \centering
    \caption{Locations of Earth-based observatories}
    \begin{tabular}{cc}
        \hline\hline
         Observatory Name& Location  \\ 
         \hline
         IceCube Neutrino Observatory& $89^\circ \mathrm{S},63^\circ \mathrm{E}$\\
         European Southern Observatory-Paranal&$24^\circ \mathrm{S},70^\circ \mathrm{W}$\\
         Quito Astronomical Observatory &$0^\circ \mathrm{S},78^\circ \mathrm{W}$\\
         Purple Mountain Observatory &$32^\circ \mathrm{N},118^\circ \mathrm{E}$\\
         Helsinki Observatory
         &$60^\circ \mathrm{N},25^\circ \mathrm{E}$\\
         \hline
    \end{tabular}
    \label{tab:2}
\end{table}

\begin{figure}[htbp]
    \centering \includegraphics[width=8.5cm,height=7.2cm]{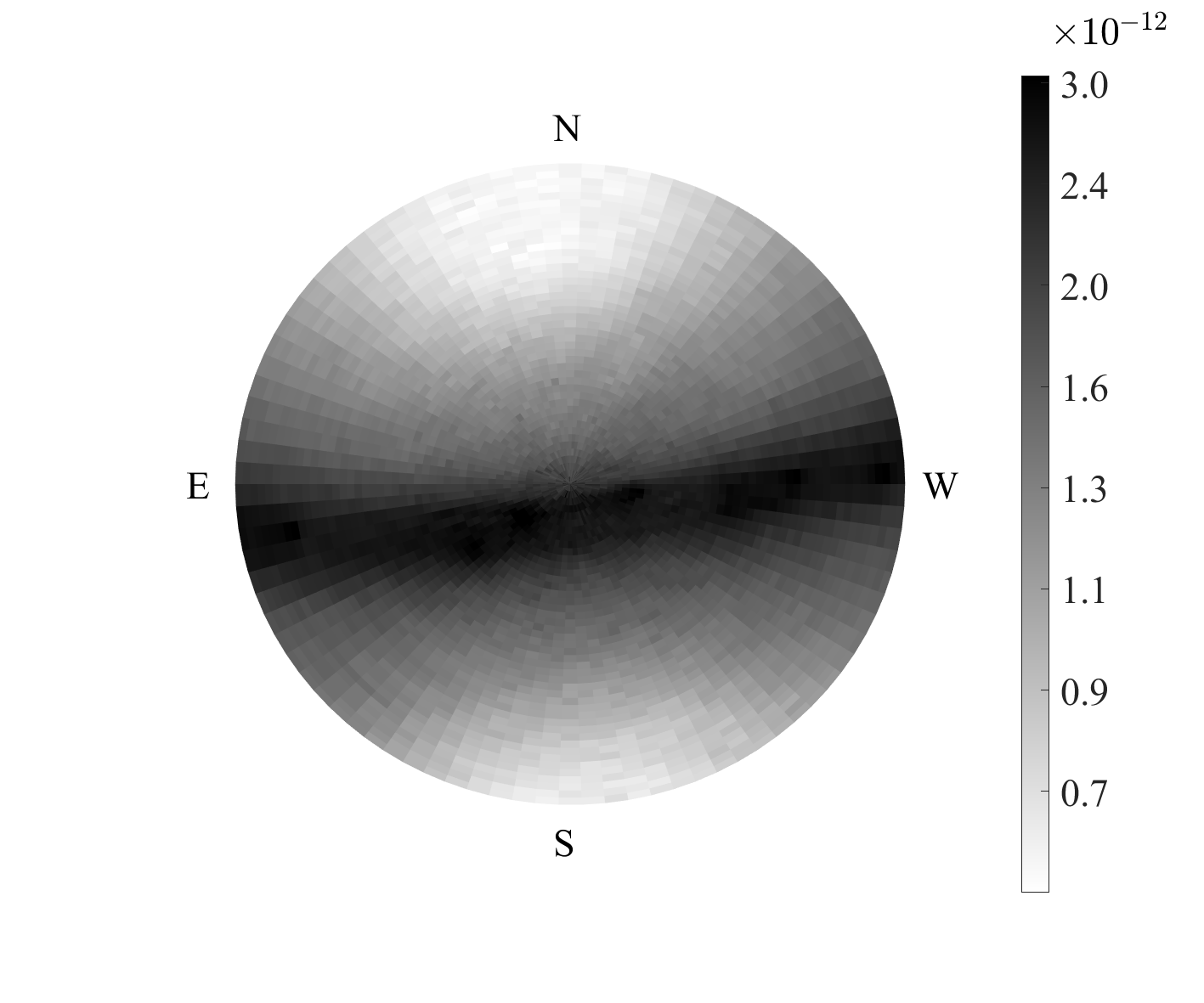}
    \caption{Fraction of solid angle subtended by particles per angular bin from the perspective of the Purple Mountain Observatory, at local midnight. The intervals of the azimuth and the elevation are $2^\circ$ and $4^\circ$, respectively. The outer circle denotes the local horizon. }
    \label{fig:Purple}
\end{figure}

\begin{figure}
    \centering \includegraphics[width=8.5cm,height=7.2cm]{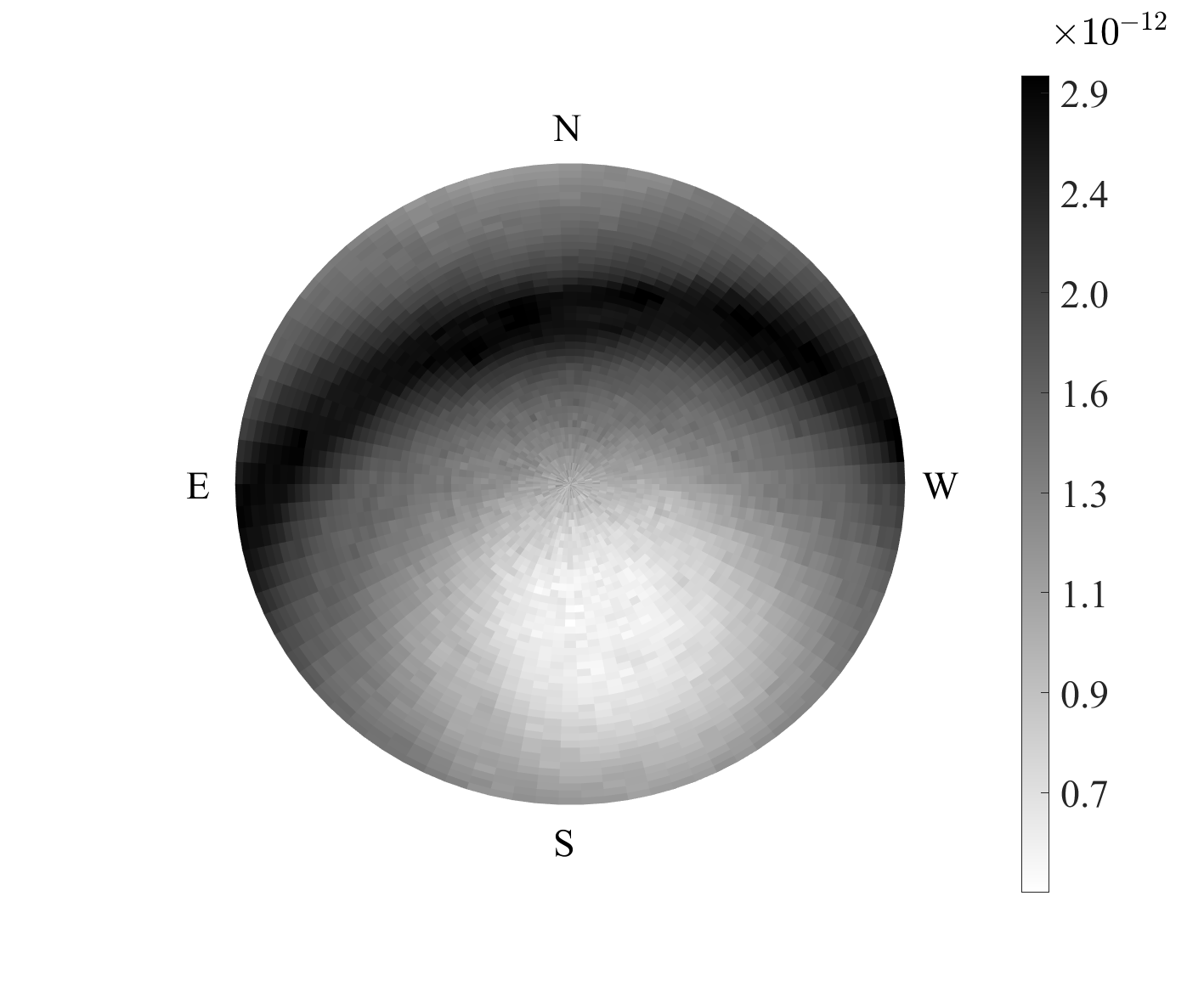}
    \caption{Fraction of solid angle subtended by particles per angular bin from the perspective of the Paranal Observatory of the European Southern Observatory, at local midnight.}
    \label{fig:EuropeSouth}
\end{figure}

\begin{figure}
    \centering \includegraphics[width=8.5cm,height=7.2cm]{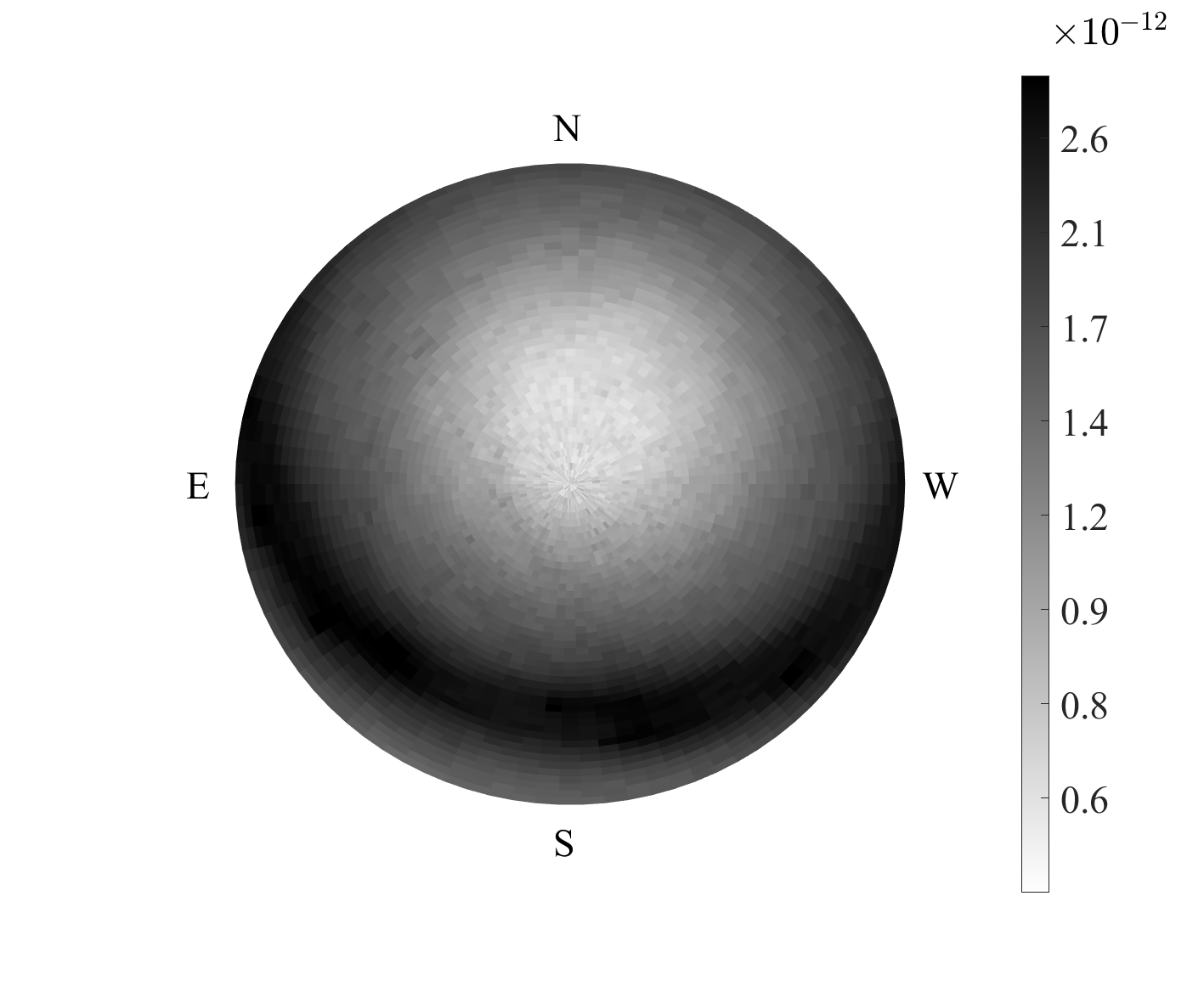}
    \caption{Fraction of solid angle subtended by particles per angular bin from the perspective of the IceCube Neutrino Observatory, at local midnight.} 
    \label{fig:IceCube}
\end{figure}

\begin{figure}
    \centering \includegraphics[width=8.5cm,height=7.2cm]{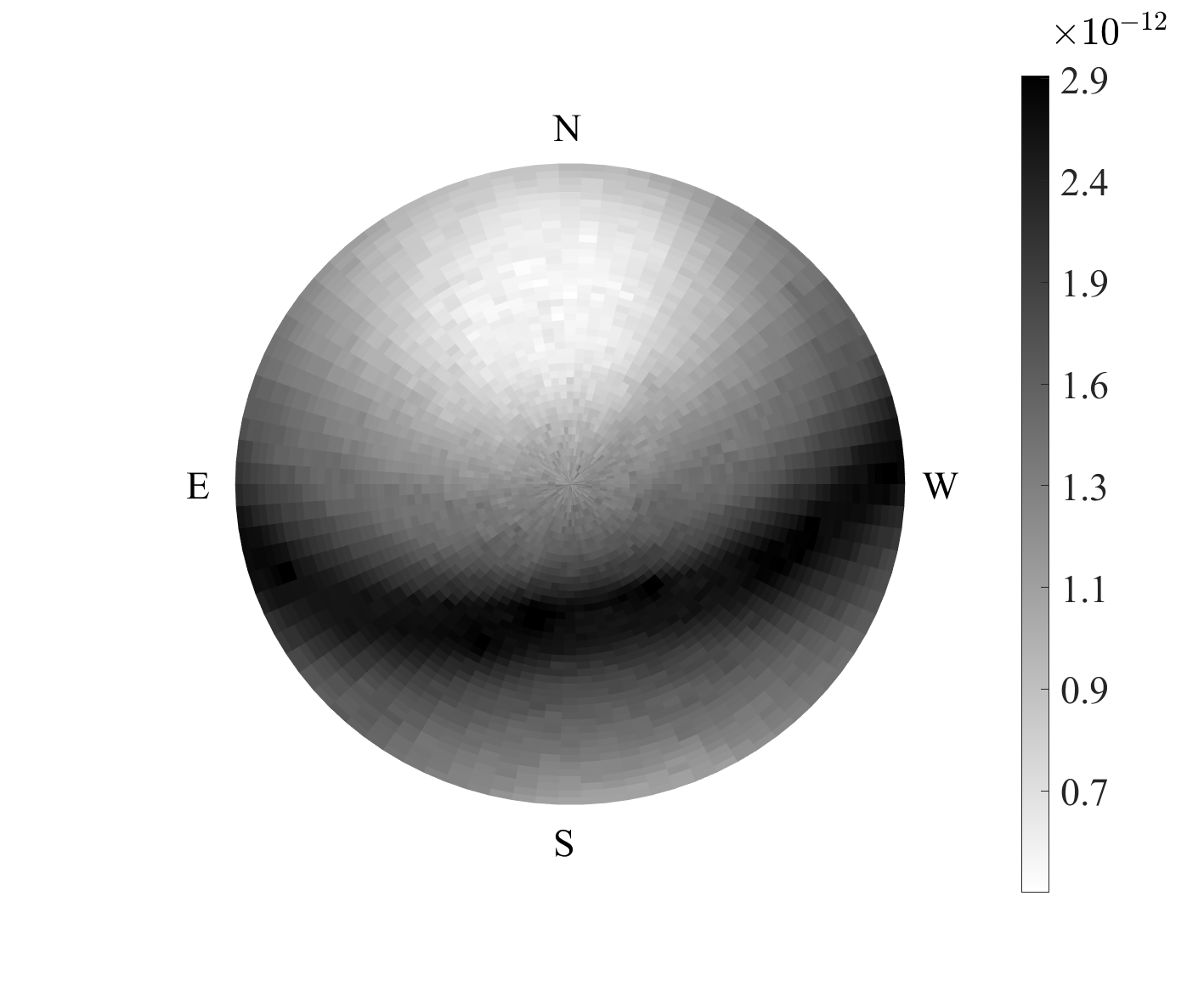}
    \caption{Fraction of solid angle subtended by particles per angular bin from the perspective of the Helsinki Astronomical Observatory, at local midnight.}
    \label{fig:Helsinki}
\end{figure}

\begin{figure}
    \centering \includegraphics[width=8.5cm,height=7.2cm]{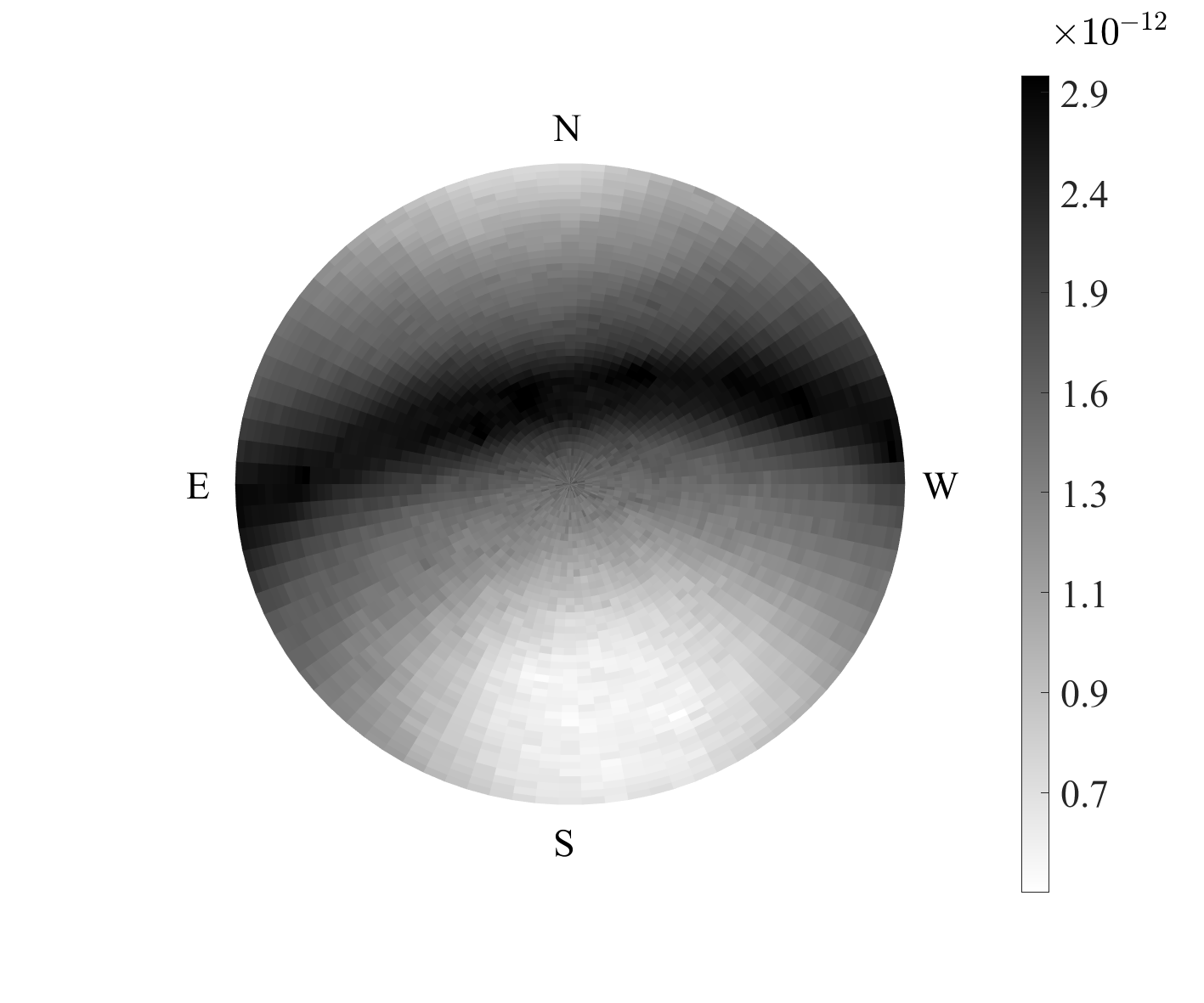}
    \caption{Fraction of solid angle subtended by particles per angular bin from the perspective of the Quito Astronomical Observatory, at local midnight.}
    \label{fig:Quito}
\end{figure}

Figs.~\ref{fig:Purple}-\ref{fig:Quito} present the fraction of the solid angle subtended by particles per angular bin from the perspectives of different observatories, where the axes are measured in terms of local azimuth and elevation. From Fig.~\ref{fig:Purple}, lunar-ejected particles form a dust belt stretching in the east-west direction, and hardly affect the observation of targets in the northern and southern sky. When observed from the Paranal Observatory of the European Southern Observatory, lunar-ejected particles form an arc in the northern half of the sky, from about $20^\circ$ south of east to about the due west (see Fig.~\ref{fig:EuropeSouth}). 

To show the influence of the lunar-ejected dust on the polar regions of the Earth, particles are projected into the sky of the IceCube Neutrino Observatory, which is pretty close to the south pole. From this perspective, particles form an arc 
near the local horizon in the southeastern direction, while the northwestern sky is less dusty (Fig.~\ref{fig:IceCube}). Since the north pole is primarily covered by the Arctic Ocean, particles are projected to the sky of the Helsinki Observatory (a typical high-latitude observatory located in the northern hemisphere of the Earth), as shown in Fig.~\ref{fig:Helsinki}. Particles are more likely to be observed in the southern sky, forming an arc from the due east to about $20^\circ$ north of west. When observing from the near-equator Quito Astronomical Observatory, observing targets in the southern sky are hardly affected by these lunar-ejected dust particles (Fig.~\ref{fig:Quito}). 

\begin{figure}
    \centering
    \subfigure[]{   \includegraphics[width=0.45\textwidth]{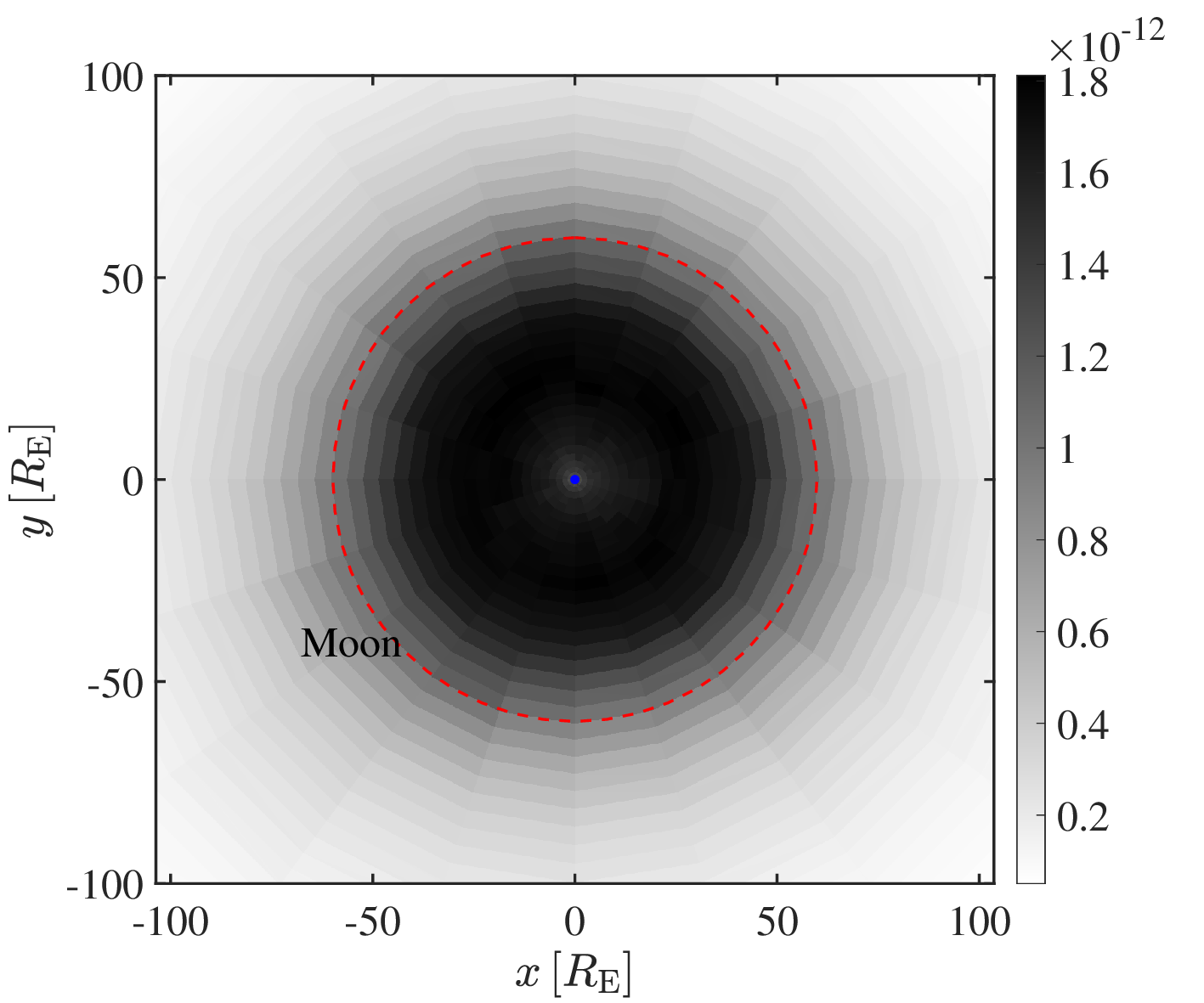}
    \label{opt}
    }
    \subfigure[]{   \includegraphics[width=0.45\textwidth]{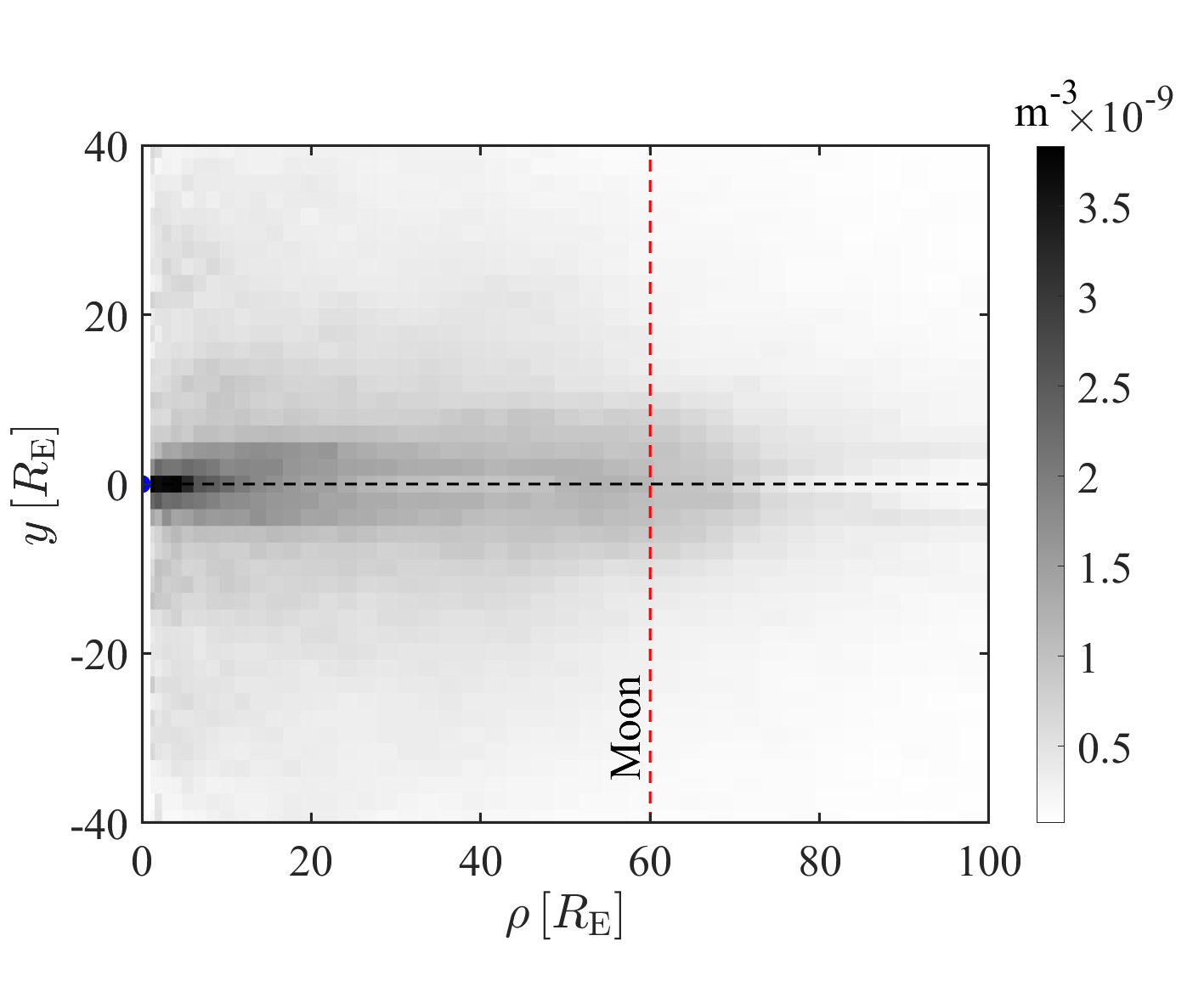}
    \label{density}
    }
    \caption{Geometric Optical depth and number density of lunar-ejected particles relative to the Earth-Moon orbital plane, where the $x$ axis points to the ascending node of the Moon at the J2000 epoch, the $z$ axis is along the normal of the orbital plane (north), and the $y$ axis is determined by the right-handed rule. \emph{Panel a}: Normal geometric optical depth. \emph{Panel b}: Azimuth-averaged number density in the $\rho$-$z$ plane, where $\rho=\sqrt{x^2+y^2}$. The blue circle denotes the Earth, and the red line denotes the orbit of the Moon.}
    \label{fig:numberdensity}
\end{figure}

\begin{figure}
    \centering \includegraphics[width=8.5cm,height=6.2cm]{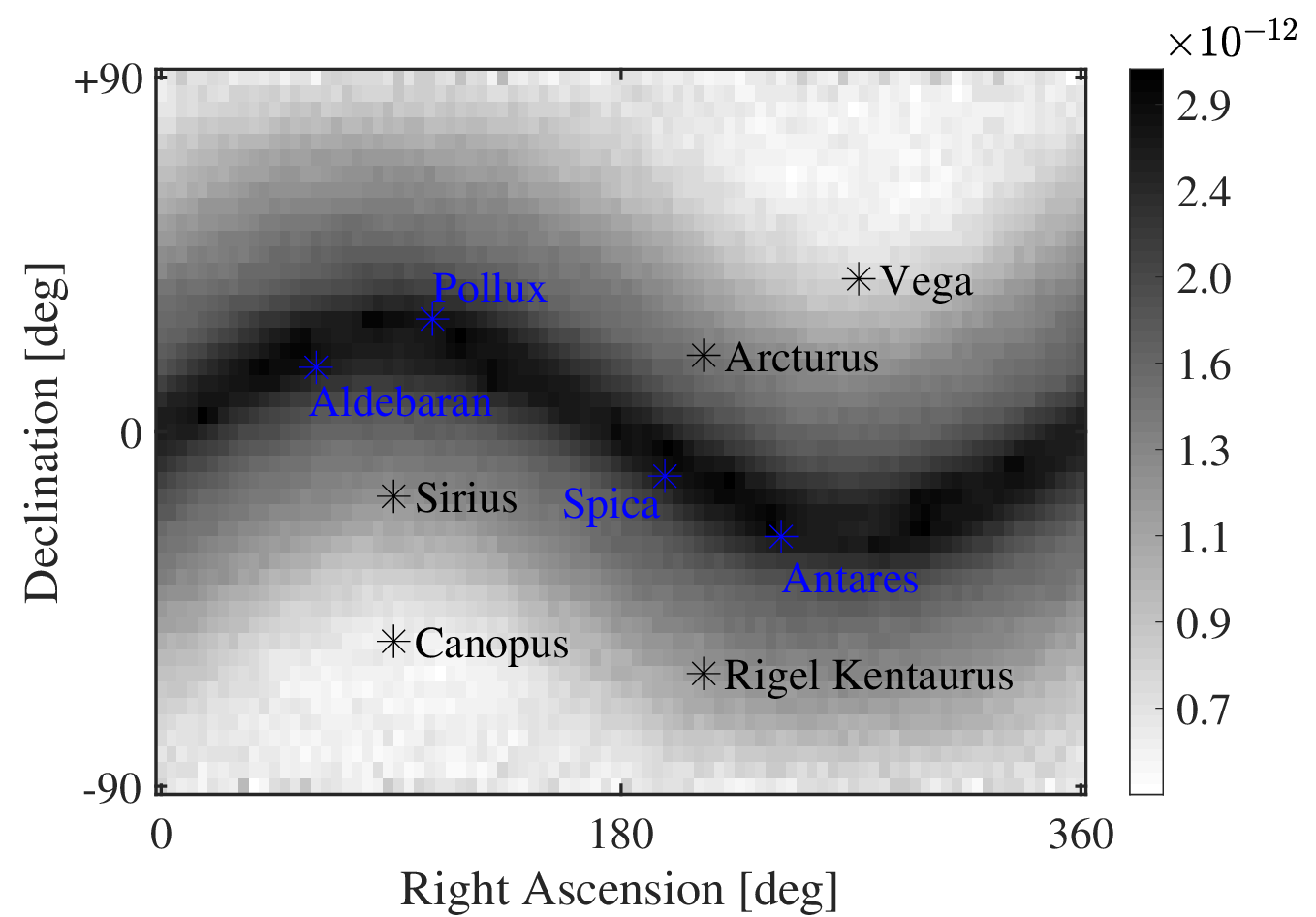}
    \caption{Fraction of solid angle subtended by particles in the ECI frame. The asterisks denote the bright stars, and the blue ones denote the stars that are more prone to be obscured by the lunar-ejected dust particles.} 
    \label{fig:LatLon}
\end{figure}

Fig.~\ref{fig:numberdensity} shows the geometric optical depth and the number density of the lunar-ejected dust particles relative to the Earth-Moon orbital plane, seen from which these particles form a dust torus around the Earth. From Fig.~\ref{density}, this dust torus is symmetric about the orbital plane, and the number density drops with the increasing radial distance from the Earth. Note that the plane of the dust torus does not coincide with the Earth's equatorial plane, so this torus may exhibit distinct configurations when observed at different observatories and local times.

Additionally, we provide the projection of lunar-ejected particles in the equatorial coordinate system, which is independent of the observation location and time (Fig.~\ref{fig:LatLon}). Several bright stars are marked in this figure, seen from which stars with low declinations, i.e., Aldebaran, Pollux, Spica, and Antares, are more prone to be obscured by lunar-ejected dust particles. 

\section{Conclusion}
In our previous work, we adopted the plausible initial distribution and production rate for lunar-ejected particles from the LDEX measurement, and implemented the long-term orbital simulation for these particles. From previous simulation results, we find about $2.3\times10^{-4}\,\mathrm{kg/s}$ lunar-ejected particles impact the Earth after long-term orbital evolution, which is a non-negligible value and somehow affect the near-Earth space environment. 

In this work, we provide the simulation results of these Earth impactors. We find that particles ejected from the lunar surface are more likely to reach the Earth if they are (1) larger than $1\,\mathrm{\mu m}$; (2) within the velocity range between $v_\mathrm{esc}$ and $1.27\,v_\mathrm{esc}$; (3) ejected in the opposite direction of the Moon's orbital motion.

We find that most of the lunar-ejected Earth impactors reach the Earth within one year, while most of the small ones ($87.2\%$ of $0.2\,\mathrm{\mu m}$ particles and $64.6\%$ of $0.5\,\mathrm{\mu m}$ particles) reach the Earth within one week, which is consistent with previous studies for small impactors \citep[e.g.,][]{alexander1984mass,yamamoto1996orbital}. We also find that a small fraction of impactors (about $30.0\%$) transfer from the Moon to the Earth after long-term orbital evolution, with the travel time of more than one year. Although these particles only account for a small proportion, they make a great contribution to the dust torus between the Earth and Moon reported by \cite{yang2022}. Besides, a considerable proportion of lunar-ejected Earth impactors can be distinguished from the interplanetary dust particles according to the differences in orbital distributions.

Particles are projected to several Earth-based observatories thus the dust distributions from the perspective of different observatories are obtained. For example, particles are most likely to be observed in the northern sky of observatories located in the southern hemisphere of the Earth (e.g.,~the Paranal Observatory of the European Southern Observatory) and in the opposite direction for the observatories located in the northern hemisphere with high latitude (e.g.,~the Helsinki Astronomical Observatory). Particles occupy an area in the northern sky of equatorial observatories near the zenith, while they form an arc near the local horizon when observed from the perspective of southern polar regions.  

Inevitably, our simulation results are subject to some uncertainties, mainly due to the mass production and the initial distribution of parameters we adopt. In such case, the uncertainties of the fraction of the solid angle subtended by particles and the number density are at least one order of magnitude. However, our result and conclusion are still useful for finding the suitable range of initial parameters for Earth impactors, predicting the visible location and orientation for lunar-ejected particles, and analyzing their effect on Earth-based observation.

\begin{acknowledgements}
     This work was supported by the National Natural Science Foundation of China (No.~12311530055, 12002397 and 62388101), the National Key R\&D Program of China (No.~2020YFC2201202 and 2020YFC2201101), and by the Shenzhen Science and Technology Program (Grant No.~ZDSYS20210623091808026). 
\end{acknowledgements}


\newpage
\bibliography{ref.bib}
\bibliographystyle{aa}

\end{document}